\documentclass[lettersize,journal]{IEEEtran}
\usepackage{cite}
\usepackage{amsmath,amssymb,amsfonts}
\usepackage{algorithmic}
\usepackage{graphicx}
\usepackage{booktabs}
\usepackage{array}
\usepackage{textcomp}
\usepackage{stfloats}
\usepackage[hidelinks]{hyperref}

\def\BibTeX{{\rm B\kern-.05em{\sc i\kern-.025em b}\kern-.08em
T\kern-.1667em\lower.7ex\hbox{E}\kern-.125emX}}
    
\hypersetup{
	pdftitle={SNAP-V: A RISC-V SoC with Configurable Neuromorphic Acceleration for Small-Scale Spiking Neural Networks},
	pdfauthor={K.H. Gunawardana, M.S. Peeris, H.M.W.K.G. Rambukwella, T.D.B. Wanduragala, S. Jameel, Roshan Ragel, Isuru Nawinne },
	pdfkeywords={ASIC design, edge computing, FPGA implementation, neuromorphic computing, Network-on-Chip, on-chip learning, RISC-V, Spiking Neural Networks, STDP}
}

\newcommand\CerebraSTotPow{518.01}
\newcommand\CerebraHTotPow{500.10}
\newcommand\CerebraSMaxFreq{10.17}
\newcommand\CerebraHMaxFreq{96.24}
\newcommand\MaxFreqImprovFactor{9.46}

\begin{document}

\title{SNAP-V: A RISC-V SoC with Configurable Neuromorphic Acceleration for Small-Scale Spiking Neural Networks}


\author{Kanishka Gunawardana, Sanka Peeris, Kavishka Rambukwella, Thamish Wanduragala, Saadia Jameel, \\Roshan Ragel, and Isuru Nawinne

\thanks{This work was supported by the Department of Computer Engineering, Faculty of Engineering, University of Peradeniya, Sri Lanka.}%
\thanks{All authors are with the Department of Computer Engineering, Faculty of Engineering, University of Peradeniya, Peradeniya 20400, Sri Lanka (e-mail: \{e19129, e19275, e19309, e18379, e18147, roshanr, isurunawinne\}@eng.pdn.ac.lk).}%
\thanks{Corresponding author: M.S. Peeris (e-mail: e19275@eng.pdn.ac.lk).}}


\maketitle

\begin{abstract}
Spiking Neural Networks (SNNs) have gained significant attention in edge computing due to their low power consumption and computational efficiency. However, existing implementations either use conventional System on Chip (SoC) architectures that suffer from memory–processor bottlenecks, or large-scale neuromorphic hardware that is inefficient and wasteful for small-scale SNN applications. This work presents SNAP-V, a RISC-V–based neuromorphic SoC with two accelerator variants: Cerebra-S (bus-based) and Cerebra-H (Network-on-Chip (NoC)-based) which are optimized for small-scale SNN inference, integrating a RISC-V core for management tasks, with both accelerators featuring parallel processing nodes and distributed memory. Experimental results show close agreement between software and hardware inference, with an average accuracy deviation of 2.62\% across multiple network configurations, and an average synaptic energy of 1.05\,pJ per synaptic operation (SOP) in 45\,nm CMOS technology. These results show that the proposed solution enables accurate, energy-efficient SNN inference suitable for real-time edge applications.
\end{abstract}


\section{Introduction}
\label{sec:introduction}

\IEEEPARstart{A}{rtificial} Intelligence (AI) has made remarkable progress across diverse application domains, starting from fundamental architectures such as Multi Layer Perceptrons (MLPs) and Artificial Neural Networks (ANNs) to more advanced Deep Neural Networks (DNNs). These networks have enabled breakthroughs in areas ranging from image recognition and natural language processing to autonomous systems. However, despite their success, DNNs are inherently resource intensive, demanding significant computational power, energy, and memory bandwidth. Such requirements limit their deployment in real time and energy constrained environments, especially in embedded and edge computing applications~\cite{yamazaki_spiking_2022}. To overcome these limitations, SNNs have emerged as a biologically inspired alternative, offering event driven computation and temporal information processing capabilities.

SNNs communicate using discrete spikes, closely mimicking the operation of biological neurons. A neuron generates a spike when its membrane potential, influenced by incoming signals, exceeds a threshold, after which it resets. This mechanism ensures sparse and asynchronous neuronal activity, enabling high energy efficiency and low computational overhead. Furthermore, the inherent temporal dynamics of SNNs make them suitable for processing time dependent data and real world sensory information. Owing to these characteristics, small scale SNNs have shown remarkable potential in various domains that demand low power and real time responsiveness. A notable example is NeuroPod~\cite{gutierrez-galan_neuropod_2020}, which employs only 22 neurons for insect-inspired locomotion. However, these kinds of compact networks are frequently deployed on massively over-provisioned neuromorphic platforms. In the case of Neuropod, it is the SpiNNaker~\cite{furber_spinnaker_2014} platform which can support up to 1 billion neurons. This mismatch results in less than 0.000002\% resource utilization, leaving the vast majority of the hardware idle while still consuming static power.


This mismatch is symptomatic of a broader structural tension in the neuromorphic computing landscape. Platforms such as Loihi~\cite{davies_loihi_2018}, Darwin3~\cite{ma_darwin3_2024}, and SpiNNaker~\cite{furber_spinnaker_2014} were architected to overcome the Von Neumann bottleneck by integrating massively parallel, event-driven, spike-based computation directly into specialized silicon substrates, eliminating the costly data transfers between processing units and memory that plague traditional architectures. However, this design philosophy targets large-scale neural workloads, and as a consequence these systems are inherently complex, costly, and power-hungry when applied to small-scale or embedded applications. The memory infrastructure required to store millions of synaptic weights and neuron states dominates the power budget~\cite{gunawardanaNeuromorphicArchitecturesEdgeOriented2026}.

Moreover, existing neuromorphic architectures often lack the flexibility required for integrating multiple computational tasks within a single platform. They are typically specialized for spike-based computation alone, limiting their ability to handle auxiliary functions such as sensor interfacing, control, and signal preprocessing, which are essential for embedded and robotic systems~\cite{gunawardanaNeuromorphicArchitecturesEdgeOriented2026}. For example, \cite{9651607} require a Raspberry Pi host alongside their $\mu$Caspian~\cite{mitchell_small_2020} neuromorphic core to handle training and environmental simulation, explicitly proposing future integration to address this limitation. Additionally, the high cost and limited accessibility of research-grade platforms such as Loihi pose significant barriers for academic institutions seeking to prototype SNN algorithms. These factors motivate the development of a new class of small scale, configurable, and cost effective neuromorphic SoCs suitable for both practical edge deployments and accessible research platforms.

In this context, this paper aims to design and implement a System on Chip (SoC) featuring a neuromorphic accelerator specifically optimized for small scale SNN applications. The proposed architecture integrates the accelerator within a RISC-V based SoC framework, enabling interaction between general purpose processing and neuromorphic computation. Leveraging the modular and open source nature of RISC-V, the system facilitates efficient management of SNN workloads while supporting additional embedded tasks such as sensor processing, control logic, and signal encoding within a unified platform, eliminating the need for external microcontrollers or companion chips.

The overarching goal of this research is to bridge the gap between large scale neuromorphic hardware and practical, small scale embedded systems. The proposed SNAP-V: \textbf{S}piking \textbf{N}eural \textbf{A}ccelerated \textbf{P}rocessor, RISC-\textbf{V} SoC with configurable neuromorphic acceleration provides a resource efficient solution capable of executing small scale SNN workloads with improved energy efficiency. By enabling  flexible interfacing and general purpose computation within a unified architecture, the design aims to address potential deployment scenarios and practical applications in robotics, edge intelligence, and low power adaptive control.

The main contributions of this work can be summarized as follows:
\begin{enumerate}
    \item The design and implementation of a RISC-V integrated neuromorphic SoC, optimized for inference on small-scale SNNs.
    \item A neuromorphic accelerator with two progressive designs: Cerebra-S (Small) establishing a baseline for performance analysis and optimization insights, and Cerebra-H (High-performance), a clustered architecture with hierarchical NoC, refined based on Cerebra-S findings.
    \item RTL-level functional validation demonstrating close agreement between software and hardware inference, with an average accuracy deviation below 3\%.
    \item A detailed power and energy characterization revealing memory-dominated system behavior and achieving low synaptic energy per operation.
\end{enumerate}

\section{Background Study}
\label{sec:background}

\subsection{Spiking Neural Networks (SNNs)}
\label{subsec:snns}

SNNs represent the third generation of neural network models, extending conventional artificial networks by incorporating temporal dynamics and event-driven computation. Unlike traditional ANNs, where neurons communicate through continuous activations, SNNs transmit discrete spike events, enabling sparse and asynchronous processing. This characteristic leads to substantially lower energy consumption and improved temporal precision, which are particularly advantageous for edge and embedded systems.

At the computational level, a range of neuron models has been proposed to emulate biological spiking behavior. Highly detailed models such as Hodgkin–Huxley~\cite{hodgkin_quantitative_1952} provide strong biological realism but are computationally intensive and impractical for hardware implementation. Simplified models, most notably the Leaky Integrate-and-Fire (LIF) neuron~\cite{abbott_lapicques_1999}, abstract spiking dynamics using lightweight equations that significantly reduce computational and hardware overhead. More complex formulations such as Izhikevich~\cite{izhikevich_simple_2003} and Resonate-and-Fire neurons~\cite{shresthaEfficientVideoAudio2024} offer richer firing dynamics but introduce additional control and arithmetic complexity that is often unnecessary for embedded inference tasks.

Information in SNNs is encoded as sequences of spikes using neural coding schemes that directly influence computational efficiency. Common strategies include rate coding, where firing frequency represents signal strength, and temporal codes such as time-to-first-spike (TTFS) and phase coding, which exploit precise spike timing for low-latency computation. The choice of neuron model and coding scheme strongly impacts implementation complexity, power consumption, and latency in neuromorphic hardware~\cite{guoNeuralCodingSpiking2021}. In this work, the LIF model is selected due to its deterministic timing behavior and hardware simplicity, making it well suited for compact neuromorphic architectures targeting real-time embedded operation.

\subsection{Demand for Small-Scale SNNs}
\label{subsec:small_scale_snns}

While large-scale SNNs aim to replicate complex cortical networks, many embedded applications operate efficiently on compact, task-specific architectures. In this context, small-scale SNNs can be defined as networks with approximately 10 to 2000 neurons. Networks of this size are commonly used in robotics, signal processing, and sensor-driven applications, where energy efficiency, deterministic behavior, and low latency are critical.

Despite the availability of large-scale neuromorphic platforms such as Intel's Loihi (131,072 neurons)~\cite{davies_loihi_2018} and SpiNNaker (up to 1 billion neurons in multi-chip configurations)~\cite{furber_spinnaker_2014}, many real-world applications require significantly smaller networks, leading to substantial hardware underutilization. In robotics, Gridbot~\cite{tang_gridbot_2018} employs only 1,321 neurons for spatial navigation, while NeuroPod~\cite{gutierrez-galan_neuropod_2020} uses merely 22 neurons on the massive SpiNNaker platform for insect-inspired locomotion. Control applications demonstrate even more extreme mismatches: autonomous racing systems~\cite{patton_neuromorphic_2021} operate with 39 neurons, lane-keeping controllers~\cite{bing_end_2018} require 34 neurons, and event-based PID control for quadrotors~\cite{stagsted_event-based_2020,dupeyroux_neuromorphic_2021} utilizes approximately 35-40 neurons. Beyond control tasks, robot localization systems~\cite{hines_compact_2024} typically employ 700--800 neurons, while micro-scale engine optimization~\cite{schuman_real-time_2021} operates with as few as 8 neurons. These examples show that small-scale SNNs are not simplified versions of large networks but domain-optimized architectures tailored for specific spatiotemporal tasks with predictable latency, efficient memory access, and minimal communication overhead, characteristics that would be better served by dedicated small-scale neuromorphic platforms rather than running on massively over-provisioned hardware.


\subsection{Neuromorphic Hardware Landscape}
\label{sec:neuromorphic_hardware}

Neuromorphic computing systems employ specialized hardware to emulate the event-driven and parallel characteristics of biological neural processing. Most existing implementations, whether ASIC or FPGA based, have been developed as standalone accelerators targeting large-scale neural simulations. While these designs achieve impressive scalability and power efficiency, they are often application-specific and lack integration with general-purpose computing or peripheral control. Consequently, limited attention has been given to compact, small-scale architectures optimized for embedded, real-time, and low-power neuromorphic computation.

Among ASIC-based platforms, several large-scale architectures have established key milestones in neuromorphic design. Intel’s Loihi integrates 128 neuromorphic cores, each simulating up to 1,024 neurons, while its successor, Loihi 2, expands this capacity to 8,192 neurons per core, offering advanced configurability and increased network scale~\cite{davies_loihi_2018,orchard_efficient_2021}. IBM’s TrueNorth achieves ultra-low-power operation of only 65 mW while simulating 1 million neurons and 256 million synapses~\cite{akopyan_truenorth_2015}. Likewise, SpiNNaker scales to millions of neurons using massively parallel ARM cores~\cite{furber_spinnaker_2014}. Complementing these systems, Darwin NPU~\cite{ma_darwin_2017} provides between 2,048 and 32,768 configurable neurons using eight physical time-multiplexed cores. Darwin3~\cite{ma_darwin3_2024} supports up to 2.35 million neurons, with on-chip learning enabled through a domain-specific instruction set architecture and flexible neuron-model programming within a 24×24 mesh network that incorporates synaptic compression for memory efficiency. The Speck processor~\cite{synsenseSpeckEventDrivenNeuromorphic2022} integrates 328,000 neurons across nine SNN cores and employs an asynchronous architecture that achieves sub-milliwatt power consumption with support for 1,024 fan-in and fan-out connections. Together, these architectures demonstrate remarkable scalability but are designed primarily as standalone accelerators for research and high-performance simulations, making them less suitable for embedded deployment where area, power, and interface constraints dominate.

More compact ASIC designs have also been proposed, focusing on efficiency rather than scale. The ODIN processor~\cite{frenkel_0086-mm2_2019}, with 256 neurons in a 0.086 mm$^2$ area, demonstrates high-density spike-based computation at extremely low energy cost. Similarly, a 4096-neuron SNN chip fabricated using 10 nm CMOS~\cite{chen_4096-neuron_2019} achieves 3.8 pJ per synaptic operation, emphasizing power efficiency and on-chip configurability. The DYNAPs processor~\cite{moradi_scalable_2018} offers a 1,024-neuron mixed-signal architecture with no on-chip learning and energy costs in the tens of picojoules per synaptic operation, emphasizing efficient inter-core routing and scalable memory. Other small-scale prototypes, such as hierarchical NoC-based and time-multiplexed neuron architectures~\cite{fang_scalable_2018,carrillo_scalable_2013}, explore resource-efficient computation within a few hundred to few thousand neurons. However, these designs generally operate as isolated accelerators. This results in limited flexibility for integration with broader system functions such as sensing or control.

FPGA-based implementations complement these ASIC efforts through their reconfigurability and rapid prototyping capabilities. Systems like SpikeHard~\cite{clair_spikehard_2023} demonstrate large-scale emulation with over 250,000 neurons, achieving high performance in parallel event-driven tasks. On the smaller end, the \textmu Caspian platform~\cite{mitchell_small_2020,mitchell_caspian_2020} supports 256 neurons and 4,096 synapses while consuming around 10–20 mW, making it suitable for embedded experimentation. Similarly, simplified FPGA-based SNN designs with a few hundred neurons~\cite{gupta_fpga_2020} achieve fast and energy-efficient inference using lightweight spiking models. Recent designs such as SYNtzulu~\cite{leone_syntzulu_2025} and SeaSNN~\cite{geng_hardware_2025} further illustrate this trend, offering efficient feed-forward inference without on-chip learning achieving milliwatt-level power on compact FPGAs but with limited flexibility due to their fixed, application-oriented architectures. While such systems demonstrate promise for low-power neuromorphic control, they generally function as peripheral co-processors that rely on host processors for data handling and interfacing.

In summary, existing neuromorphic hardware spans a spectrum from large-scale, power-efficient ASICs to flexible FPGA-based prototypes. Few efforts address the need for small-scale, configurable neuromorphic architectures that can coexist with general-purpose processing within a unified SoC environment. Bridging this gap is critical for realizing low-power neuromorphic computing at the edge, where real-time responsiveness, adaptability, and compact integration are essential.

\subsection{RISC-V and On-Chip Communication for Neuromorphic SoCs}
\label{subsec:riscv_noc}

RISC-V’s open-source, modular ISA provides an ideal foundation for the design of neuromorphic SoC architectures. Its extensible instruction set, customizable privilege modes, and lightweight core \cite{noauthor_brian_nodate, waterman_volume_nodate} implementations make it highly suitable for integrating SNN accelerators alongside general-purpose processors. Compared to proprietary ISAs, RISC-V offers unrestricted design flexibility, enabling the incorporation of neuromorphic extensions, custom spike-handling instructions, and energy-efficient data pathways without licensing constraints. Although RISC-V has been widely adopted in AI/ML, IoT, and edge SoCs, such as GAP-8~\cite{flamand_gap-8_2018} and MARSELLUS~\cite{conti_marsellus_2024}, its use in neuromorphic computing remains limited. Only a few implementations, including SpikeHard~\cite{clair_spikehard_2023} and Urgese et al.~\cite{urgese_interfacing_nodate}, have demonstrated RISC-V-based neuromorphic integration, emphasizing its potential to bridge the gap between programmable control and event-driven computation. Future RISC-V-based neuromorphic SoCs could leverage heterogeneous architectures, combining lightweight RISC-V cores with spiking accelerators and shared memory systems to achieve real-time, low-power inference.

A critical aspect of such architectures is efficient on-chip communication. As neuromorphic workloads scale, bus-based interconnects struggle with bandwidth limitations and latency bottlenecks. NoC architectures address these issues by enabling parallel, packet-switched communication among neuromorphic cores and RISC-V processors. Studies such as Fang et al.~\cite{fang_scalable_2018} and Carrillo et al.~\cite{carrillo_scalable_2013} demonstrate that hierarchical and mixed-mode NoC topologies achieve scalable, low-latency spike communication while minimizing silicon overhead. Hierarchical NoC (H-NoC) frameworks partition the network into local and global domains, allowing clusters of neurons to communicate asynchronously with reduced contention and improved energy efficiency. Similarly, hybrid routing schemes combining mesh and hierarchical layers optimize throughput and fault tolerance for large-scale neuromorphic arrays. These findings highlight that integrating RISC-V-based control with NoC-enabled neuromorphic cores can yield scalable, flexible, and power-efficient SoC platforms offering a promising direction for next-generation event-driven computing.

\section{SoC ARCHITECTURE OVERVIEW}
\label{sec:soc_overview}

\begin{figure}[t!]
\centering
\includegraphics[width=\columnwidth]{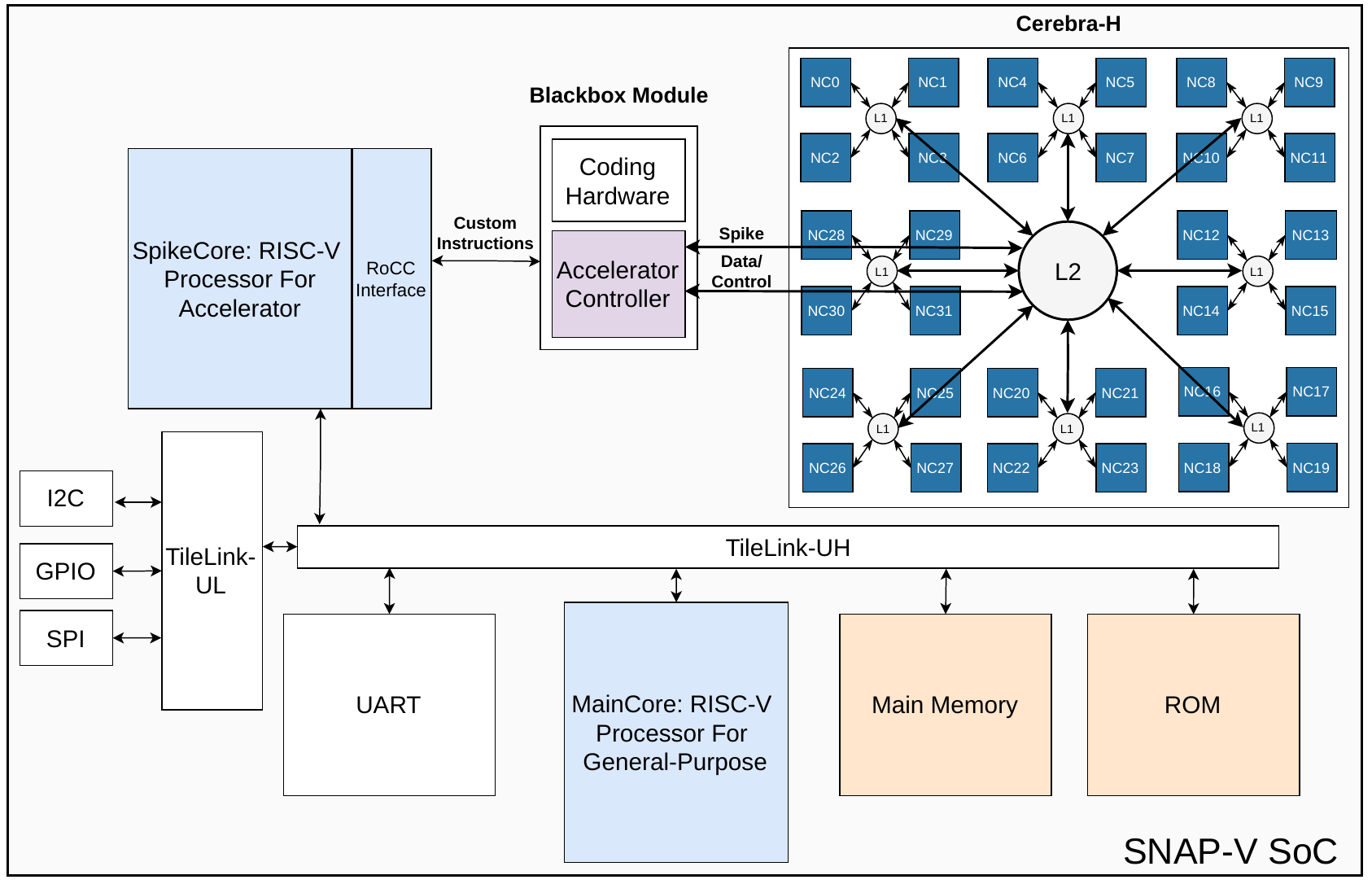}
\caption{High Level Overview of SNAP-V SOC. The architecture integrates a MainCore for general-purpose tasks and a SpikeCore for accelerator orchestration via the RoCC interface. The neuromorphic subsystem (Cerebra-H) communicates with the SoC through a dedicated Blackbox Module containing encoding/decoding hardware and an accelerator controller.}
\label{fig:snapvsoc}
\end{figure}

The SNAP-V SoC as illustrated in Fig.~\ref{fig:snapvsoc} integrates a general-purpose RISC-V processor subsystem with a configurable neuromorphic accelerator to form a unified platform for low-power, event-driven computation. The design aims to bridge the gap between digital processors and biologically inspired computing by combining programmability and real-time responsiveness within a single chip.

The SoC employs a dual-core architecture consisting of the MainCore and SpikeCore, both based on the RocketChip~\cite{noauthor_31_nodate} RISC-V implementation. The MainCore manages system control, initialization, and software workload, while the SpikeCore handles spiking network orchestration and accelerator communication. Both cores share peripheral and memory subsystems via a TileLink-based interconnect.

Peripheral components including UART, GPIO, SPI, and I2C enable real-time sensor and actuator interaction, while on-chip ROM and main memory support instruction execution and spike-data buffering. The integration of a neuromorphic accelerator, accessed through a Blackbox module, allows hardware acceleration of neuron, synapse, and spike-based computations.

\subsection{RISC-V SoC Integration}
The neuromorphic subsystem is tightly integrated with the RISC-V cores using the Rocket Custom Coprocessor (RoCC) interface. This interface provides a programmable and decoupled bridge between the SpikeCore processor and the neuromorphic accelerator. Through RoCC, custom SNN instructions are issued from SpikeCore, enabling efficient configuration, spike-data transfer, and synchronization with the accelerator's event-driven pipeline.

The Accelerator Controller implements instruction decoding, dual NoC management, and time synchronization between the CPU and the accelerator. The controller distinguishes between initialization and spike-processing contexts, managing 8-bit packets used for configuring and 11-bit packets for spike communication through independent FIFO queues. Synchronization is achieved via dedicated control signals, ensuring that each simulation step advances only after all neurons and interconnect operations have completed resulting in a dynamic timestep.


By adopting RISC-V's open and modular ISA, the SNAP-V SoC achieves design flexibility and compatibility with open-source SoC frameworks such as Chipyard~\cite{chipyard}. This integration enables unified simulation, rapid hardware generation, and co-verification of neuromorphic workload using both software and hardware co-design methodologies.

\subsection{Spike Encoding and Decoding}
In conventional neuromorphic systems, neural coding and decoding are often handled by software, introducing significant latency and CPU overhead. To address this limitation, SNAP-V introduces a dedicated Coding Hardware Unit that performs on-chip neural encoding and decoding.

The encoder converts real-world sensor data such as temperature, motion, or brightness into spike trains using hardware-implemented rate coding scheme. The unit interfaces directly with the peripheral subsystem, transforming analog or digital sensor data into spike packets that are routed to neuron clusters through the NoC.

Conversely, the decoder reconstructs spike events into numerical outputs or actuator commands, closing the perception-to-action loop. This bidirectional coding mechanism significantly reduces inference latency, offloads computational load from the RISC-V cores, and enables real-time, closed-loop neuromorphic control for embedded edge applications.

By integrating spike encoding and decoding at the hardware level, SNAP-V achieves tighter coupling between sensing, computation, and actuation, making it highly suitable for adaptive, event-driven systems.


\section{Neural Accelerator Architecture Design 1: Cerebra-S}
\label{sec:cerebra_s}


\begin{figure}[t]
    \centering
    \includegraphics[width=1\columnwidth]{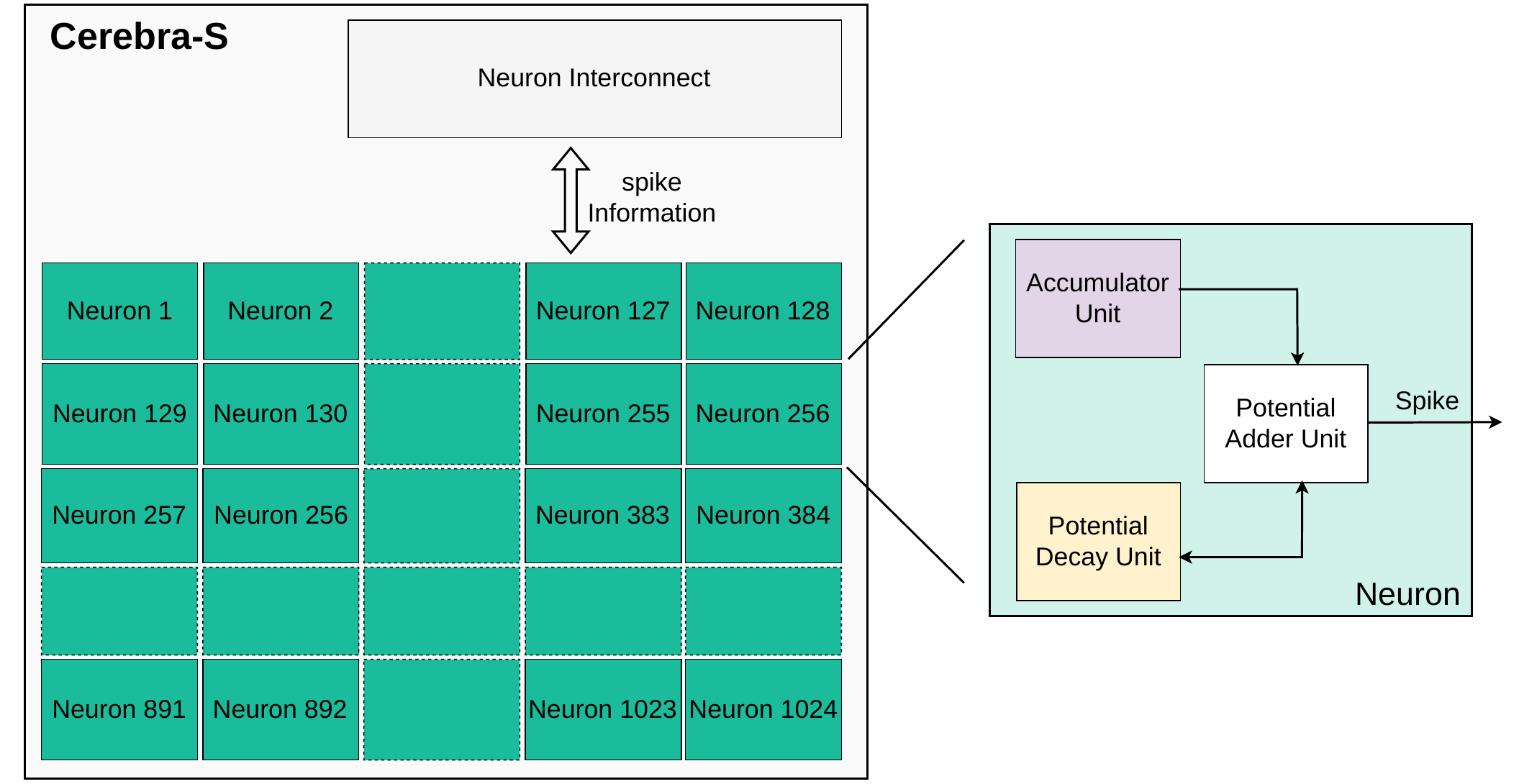}
    \caption{Accelerator Design of Cerebra-S. The architecture consists of a tiled array of 1024 physical neurons connected to a shared neuron interconnect via a global tagged bus. Each individual neuron tile contains an accumulator unit for synaptic integration, a potential decay unit, and a potential adder unit for threshold evaluation and spike generation.}
    
    \label{fig:cerebra_s_high_level}
\end{figure}

As an initial step towards the development of a neuromorphic system, a first-generation accelerator, termed Cerebra-S was designed. It intentionally adopts simple architectural design. This version serves as a feasibility study, allowing us to evaluate whether such a minimalist approach is sufficient in terms of power efficiency, performance, and suitability for deployment on edge devices. Although Cerebra-S is not optimized for performance, it establishes a baseline design upon which more optimized architectures were later developed. Insights gained from this initial design directly informed subsequent iterations that balance architectural simplicity with improved efficiency.

A high-level overview of the Cerebra-S architecture is illustrated in Fig. 2. The core consists of a fixed number of neurons, currently set to 1024, with each logical neuron mapped to a physical neuron during initialization. Each neuron comprises three functional units: accumulator unit, potential decay unit, and reset unit. The decision to use 1024 neurons was based on our literature review \cite{gunawardanaNeuromorphicArchitecturesEdgeOriented2026}, which indicated that many small-scale SNN models can be effectively supported within this size. They collectively model the firing and refractory behavior of biological neurons. The core is organized as a tiled array of neuron modules, all connected to a shared neuron interconnect via a global tagged bus. At timestep boundaries, spiking activity from the neuron array, together with externally injected stimulus spikes, is captured by the interconnect and processed using an adjacency matrix representation of synaptic connectivity stored in on-chip memory. For each spiking neuron, the interconnect sequentially traverses its outgoing synapses and emits weighted synaptic events, one per clock cycle, containing the destination neuron address and synaptic weight. These events are broadcast over the shared bus, where each neuron locally filters incoming traffic and integrates matching events into its accumulator. 

 This event-driven, adjacency matrix-based interconnect decouples spike propagation from neuron computation, enabling scalable and deterministic spike routing without dedicated per-neuron buffering.




\subsection{Neuron Interconnect}

The neuron interconnect provides the following functionality,

\begin{itemize}
    \item Housing the Neuron Placement Graph: The neuron interconnect includes a separate memory segment used to store the neuron placement graph, which stores synaptic weights of neurons within the SNN. This takes form of an adjacency matrix. The memory specifies how spikes are routed between neurons and is accessed during spike propagation to resolve communication and information passing including the weights of the connections. The matrix is housed in an SRAM, which is referenced by the interconnect when needed.
    
    \item Spike Propagation: During the inference phase, each neuron communicates with the interconnect through a
common bus interface. At each timestep the spikes will be broadcasted as one synapse
per clock cycle in the form of a tagged bus transaction
containing the destination neuron address and the synaptic
weight. The neurons will be snooping on the bus to capture their loads.


\end{itemize}

\subsection{Neuron}

\subsubsection{Accumulator Unit}

The accumulator unit serves as the entry point to each neuron and is responsible for integrating incoming spikes into the neuron's membrane potential. The accumulator accepts inputs, each consisting of a destination address and an associated 32-bit synaptic weight. When an input is received, the accumulator adds the provided weight to the neuron's accumulated potential using a fixed-point adder. Over the course of a timestep, multiple such events may be processed sequentially, resulting in an aggregated membrane potential that reflects the combined effect of all incoming spikes.

\subsubsection{Potential Decay Unit}

The Potential Decay Unit (PDU) updates the neuron's membrane potential at timestep boundaries to model temporal decay in biological neurons. In the implemented Leaky Integrate-and-Fire (LIF) model, the potential is scaled by a constant fractional decay factor, utilizing a fixed-point multiplication. To ensure accurate dynamics and eliminate unintended multi-cycle feedback delays, the PDU is designed as a purely combinational logic block without internal state registers. The resulting decayed potential is continuously forwarded to the potential adder unit for input integration, threshold comparison, and subsequent spike generation.

\subsubsection{Potential Adder Unit}

The Potential Adder Unit combines the accumulated synaptic input from the Accumulator Unit with the decayed membrane potential produced by the Potential Decay Unit. It evaluates the resulting membrane potential against a predefined threshold to determine whether the neuron spikes. Upon a spike event, the unit generates the neuron's spiking output and resets the membrane potential of the decay unit. The resulting membrane potential is forwarded for decay in the next timestep. The spike indication is captured by the neuron interconnect at the timestep boundary to trigger synaptic event generation.





\section{Neural Accelerator Architecture Design 2: Cerebra-H}
\label{sec:cerebra_h}

This section introduces \textit{Cerebra-H}, a second-generation neuromorphic accelerator derived from the Cerebra-S architecture described in Section~\ref{sec:cerebra_s}. While Cerebra-S demonstrated the viability of direct neuron-to-hardware mapping and event-driven spike propagation using a minimal global interconnect, its flat organization lacking distributed memory created a memory access bottleneck, limiting energy efficiency and performance.


Cerebra-H was made to address these limitations by adding a layer of complexity on top of the base Cerebra-S design. While every other decision in Cerebra-S favoring
simplicity could be justified, the only aspect in which the con heavily outweighed the benefits of a simple design was in communication. Having centralized memory along with a common bus interface saw the system suffering from communication and memory access bottlenecks.

The new architecture has a hierarchical routing network. Neurons are grouped into clusters that enable localized spike aggregation and weight access, while supporting parallel spike communication across clusters through the hierarchical interconnect. This organization reduces the global communication overhead, increases concurrent spike propagation, and improves timing determinism.


Total power for Cerebra-S and Cerebra-H are \CerebraSTotPow\,mW, \CerebraHTotPow\,mW respectively. The maximum clock frequency increased from \CerebraSMaxFreq\,MHz to \CerebraHMaxFreq\,MHz which is a \MaxFreqImprovFactor~times improvement. This clock frequency is of utmost value as our design is made to accommodate time-sensitive edge applications where real time response is vital.

These refinements evolve Cerebra-S from a minimal proof-of-concept design into a more structured accelerator suitable for integration with a SoC.

\begin{figure}[t!]
\centering
\includegraphics[width=0.70\linewidth]{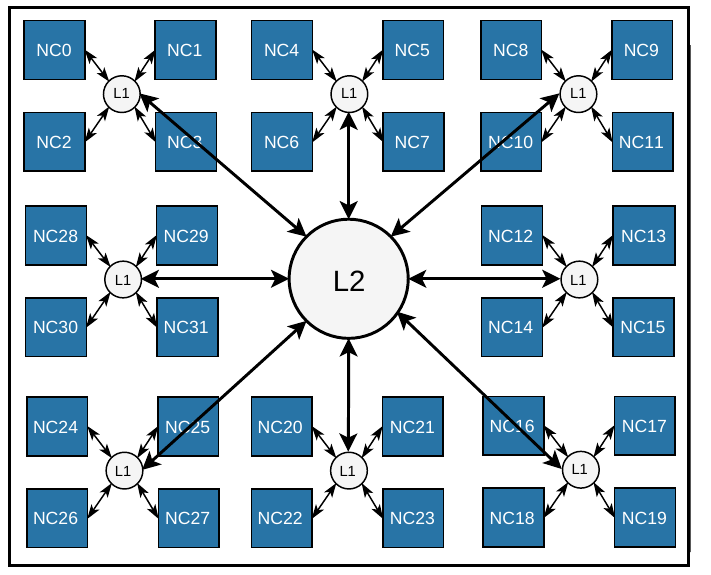}
\caption{Accelerator Design of Cerebra-H. The clustered neuromorphic architecture employs a hierarchical Network-on-Chip (NoC) topology. Lower-layer routers (L1) connect groups of four neuron clusters (NC0-NC31), which are further aggregated by a central upper-layer router (L2) to facilitate parallel spike communication and reduce global routing overhead.}
\label{fig:accelerator-design}
\end{figure}

\subsection{Enhanced Neuron Microarchitecture}
\label{subsec:enhanced_neuron}

Cerebra-H retains the same configurable Leaky Integrate-and-Fire (LIF) neuron model used in Cerebra-S while refining the internal neuron organization to improve efficiency and control. The fundamental neuron functionality remains unchanged, ensuring behavioral consistency across accelerator generations.


Each neuron incorporates a lightweight controller implemented as a finite-state machine. The controller manages parameter configuration and operational sequencing, allowing decay rate, firing threshold, and reset behavior to be programmed through configuration packets. This enables runtime adjustment of neuron behavior without modifying hardware.



Membrane potential decay is implemented using arithmetic right-shift operations to approximate leakage behavior, intentionally avoiding the computationally expensive multipliers utilized in the earlier Cerebra-S design. Decay rates of 0.125, 0.25, 0.5, and 0.75 are supported through selectable shift configurations. By replacing the fixed-point multiplication with these shift operations, Cerebra-H reduces the power consumption associated with membrane potential updates, providing necessary functional flexibility with minimal hardware overhead compared to its predecessor.


Reset behavior following a spike is configurable and supports hold (no change to the membrane potential), reset to zero, and subtractive (subtracting the threshold potential value from the membrane potential) modes, extending the neuron response options beyond those available in Cerebra-S.

\begin{figure}[t!]
\centering
\includegraphics[width=0.90\linewidth]{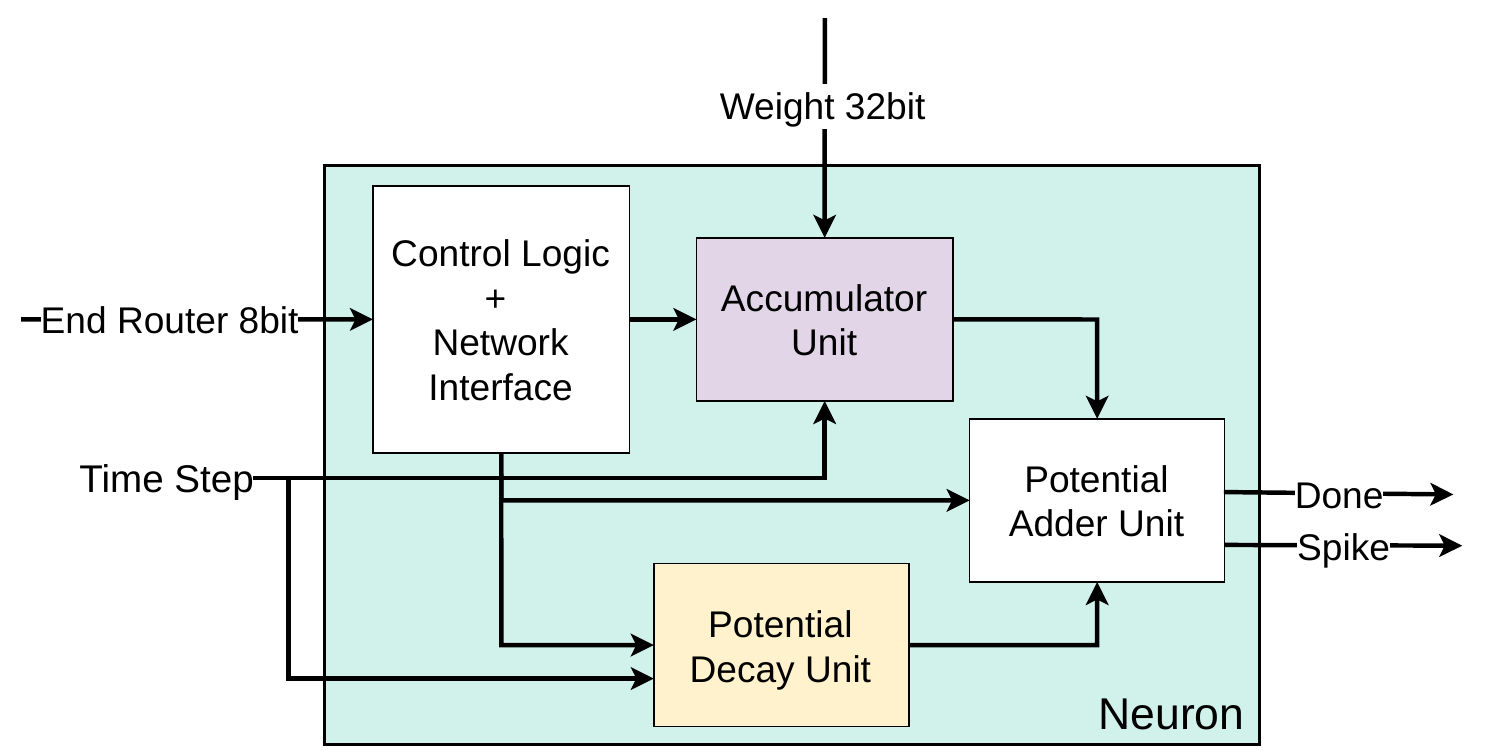}
\caption{Neuron Microarchitecture. The internal datapath of a single configurable Leaky Integrate-and-Fire (LIF) neuron. It includes a finite-state machine control logic block, an accumulator unit for integrating incoming 32-bit synaptic weights, a potential decay unit utilizing arithmetic right-shifts, and a potential adder unit that evaluates membrane thresholds to generate spike outputs.}
\label{fig:neuron-design}
\end{figure}

Unlike the global bus based connectivity in the first design, neurons in Cerebra-H interface with cluster-level forwarding and spike packet encoding logic. This change decouples neuron computation from spike routing and supports deterministic synchronization within each cluster. Fig.~\ref{fig:neuron-design} illustrates the refined neuron microarchitecture.

\subsection{Neuron Clusters}

The system is organized into clusters, each hosting 32 neurons and grouped into cluster groups that share routing and memory resources. 32 neurons hit the sweet spot where spike traffic can be comfortably communicated without excessive hardware replication. 32 such clusters make up for a total of 1024 neurons. Clustering enables us to place neurons with common synapses within the same cluster to reduce the distance spike packets should travel. Neurons with sparse connectivity between them can be placed in different clusters for minimal communication overhead. Each cluster contains a Cluster Controller, an Incoming Forwarder, and an Outgoing Encoder that interface with the NoC. The Cluster Controller sequences configuration and initialization via opcodes and manages packet-driven register updates. Supported opcodes include OPCODE\_LOAD\_NI, OPCODE\_LOAD\_IF, and OPCODE\_LOAD\_OE. Initialization packets carry structured fields (neuron row indices, flit counts, and data flits). The controller buffers and sequences multi-flit transfers to ensure integrity.

\begin{figure}[t!]
\centering
\includegraphics[width=0.9\linewidth]{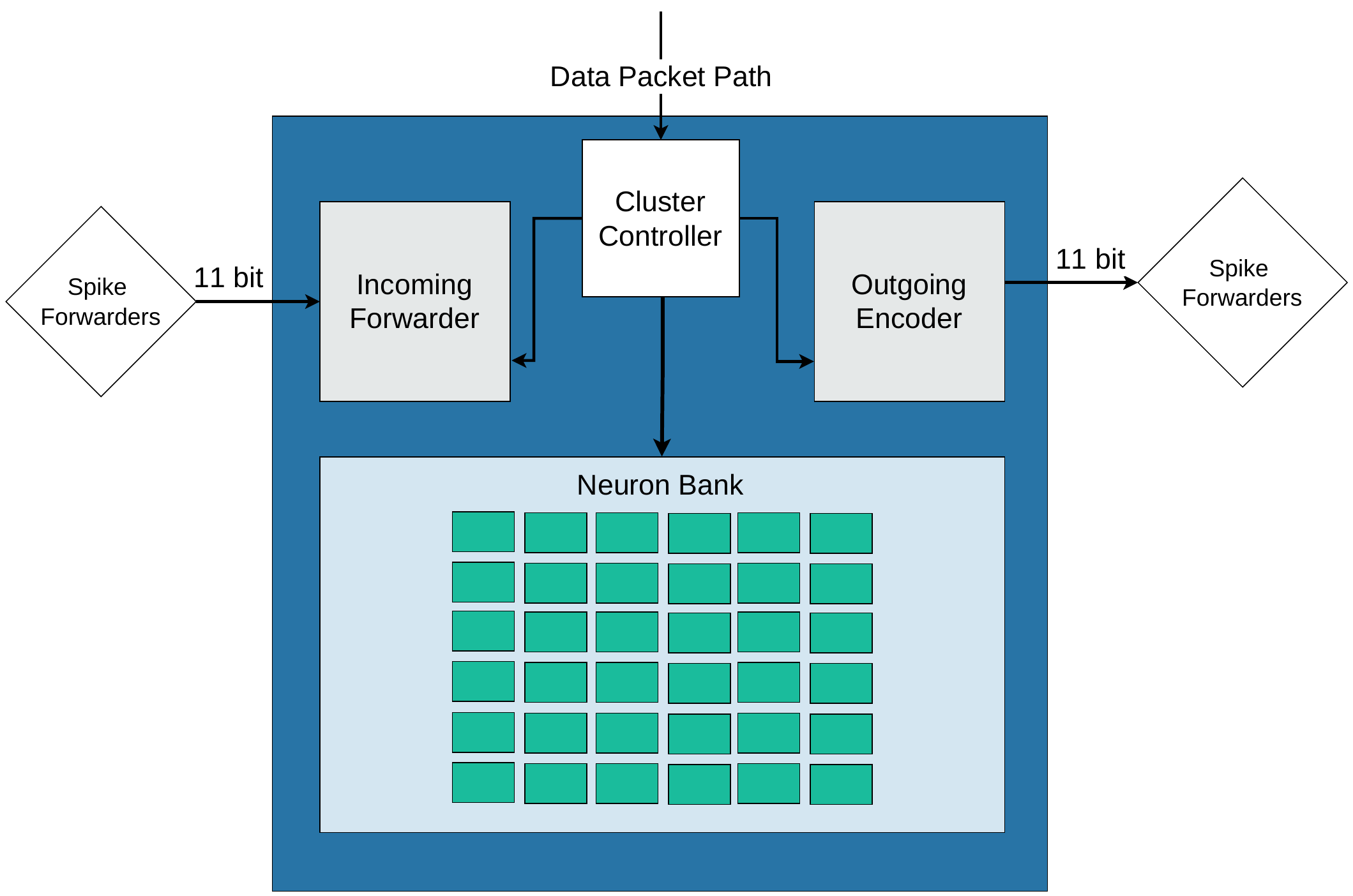}
\caption{Functional Block Diagram of a Cluster. The internal organization of a single 32-neuron cluster illustrating the dual communication paths. The cluster controller manages configuration packets, while the incoming forwarder and outgoing encoder handle the routing and serialization of 11-bit spike packets to and from the local neuron bank.}
\label{fig:cluster-architecture}
\end{figure}

As illustrated in Fig.~\ref{fig:cluster-architecture}, the cluster datapath components are as follows,
\begin{itemize}
    \item Incoming Forwarder:  Operates in two modes, 
    \begin{itemize}
        \item Forwarding mode which performs real-time lookup of incoming spike cluster and neuron IDs, generates weight addresses, fetches the corresponding weights, and delivers them to the neurons.
        \item Storage mode which programs the lookup table.
    \end{itemize}
    \item Outgoing Encoder: Serializes cluster-wide spike vectors into packets and transmits them to the level-1 router, while also monitoring congestion and ensuring that each spike is emitted exactly once per cycle.
    \item Neuron Interfaces: Each neuron connects to the Incoming Forwarder through the dedicated weight width data path and to the Outgoing Encoder through a single bit spike output line.
\end{itemize}
   
\subsection{Weight Memory}


Weight storage in Cerebra-H is organized at the cluster-group level to effectively resolve the memory access bottlenecks identified in Cerebra-S. Within this architecture, a single cluster group comprising four clusters shares a single-port weight memory for a good balance of distributedness, maximized memory usage and power efficiency. coordinated by a Weight Resolver and Incoming Forwarder per cluster. Each memory row stores a full cluster-wide weight vector whose width equals 32 times the number of neurons in a cluster, and the address space supports up to 2048 rows to manage power and area consumption. The architecture supports a total of 524,288 synaptic weights system-wide. The Incoming Forwarder mapping tables index by source cluster ID (6 bits) and source neuron ID (5 bits) to generate weight addresses, while flow-control feedback prevents resolver buffer overflow.

\subsubsection{Inference Mode (Normal Operation)}

Each cluster places read requests into its own request queue, dequeues at most one request per cycle, and issues a read to the shared memory. Because the memory read is asynchronous, the data for the granted address appears without an added cycle of latency. The resolver then marks one of four output lanes as valid to indicate which cluster is receiving the data; only that lane carries the weight row. Each request queue exposes its occupancy so that upstream logic can throttle traffic, and a completion indicator is asserted when all request queues are empty and no grant is outstanding.

\subsubsection{Initialization Mode (Setup Operation)}

The weight memory is written using a byte-serial protocol that sends the low address byte, high address byte, a byte indicating the number of data bytes to follow, and then the data bytes themselves. When the last data byte is received, the assembled row is committed to the target address. If fewer than the full row's bytes are provided, the remaining bytes default to zero. The memory also clears all rows to zero on reset. During initialization, runtime read arbitration is disabled, and reads are serviced only in normal operation. Key specifications and capabilities of the weight-resolution logic are summarized in Table~\ref{table:weight_resolver_specs}.

\begin{table}[t!]
\caption{Weight resolver specifications and capabilities}
\label{table:weight_resolver_specs}
\centering
\begin{tabular}{@{}p{0.28\linewidth}p{0.67\linewidth}@{}}
\toprule
\textbf{Specification} & \textbf{Description} \\
\midrule
Concurrent Access Ports & Four per-cluster request queues (depth eight) \\
Memory Interface & Single-port shared memory with asynchronous read; row width equals 32 times the neurons per cluster (default 1024 bits) \\
Buffer Management & Per-queue occupancy is exposed for flow control \\
Arbitration Logic & Fixed priority (port 0 over 1 over 2 over 3) with one-hot grant \\
Weight Distribution & Four parallel output lanes; only the granted lane carries data, accompanied by a per-lane valid indicator \\
Pipeline/Latency & One cycle from arbitration to a valid output lane; zero additional cycles for memory read (combinational) \\
Initialization Protocol & Byte-serial sequence: low address byte, high address byte, count of data bytes, followed by data bytes; remainder is zero-padded; memory rows are cleared on reset \\
Idle/Completion & A completion flag is asserted when all queues are empty and no grant is in flight \\
\bottomrule
\end{tabular}
\end{table}

\subsection{Spike Communication and Network-on-Chip}

To overcome the severe communication bottlenecks caused by the common bus interface of Cerebra-S, Cerebra-H introduces a hierarchical NoC architecture. As illustrated in Figure~\ref{fig:accelerator-design}: Lower-layer routers connect to exactly four neuron clusters, while upper-layer routers aggregate up to eight lower-layer routers. This arrangement reduces hop counts and simplifies routing table sizes while supporting deterministic forwarding and efficient aggregation. Additionally, if a SNN model fits within a single cluster group, spike propagation requires only the lower level router minimizing communication overhead.

The NoC is also two layered for communication separating spike traffic and control/data packet flow:
\begin{itemize}
    \item Data and Control Packet Path: Initialization is carried out in a serial uni-directional manner between the SpikeCore and Cerebra-H. Therefore, a simple buffer-free combinational router design was adopted.
    \item Spike Packet Path: Since spike communication is parallelized, a buffered and pipelined router was utilized. This design contains necessary control signals to ensure reliable completion detection and bounded resource usage.
\end{itemize}

A side-by-side detailed comparison of these two paths is provided in Table~\ref{table:dual_paths}.

\begin{table}[t!]
\caption{Comparison of dual data paths in router design}
\label{table:dual_paths}
\centering
\begin{tabular}{@{}p{0.20\columnwidth}p{0.345\columnwidth}p{0.345\columnwidth}@{}}
\toprule
\textbf{Aspect} & \textbf{Data/Control Packet Path} & \textbf{Spike Packet Path} \\
\midrule
Primary Use & System boot-up and configuration & Real-time spike communication \\
Buffering & No internal buffers & FIFO buffers at each port \\
Routing Method & Address-based routing & Predefined port mappings \\
Flow Control & Valid/ready handshake & FIFO status flags \\
Latency & Three cycles per hop & Multi-cycle pipelined \\
Traffic Pattern & Sparse, bursty & High-throughput \\
\bottomrule
\end{tabular}
\end{table}


The hierarchical NoC also enables the neuromorphic accelerator to host and execute multiple SNN models concurrently. Through packet-based communication and address-based routing isolation, different models can occupy independent neuron clusters and memory regions, allowing them to run simultaneously without interference. This design feature allows the SoC to dynamically deploy and execute multiple small-scale SNN workloads in parallel, provided that each fits within the available hardware resources, thereby improving hardware utilization and experimental flexibility.

\section{Experiments}

This section outlines the experimental methodology used to evaluate the functional correctness and hardware efficiency. The evaluation comprises two major phases: functional accuracy verification, followed by detailed power, latency, and area characterization. Together, these stages establish a comprehensive view of the accelerator's behavior, spanning from algorithmic fidelity to physical implementation.

\subsection{Functional Evaluation and Accuracy}

To validate functional behavior, the MNIST handwritten digit dataset is used as the primary benchmark. Each $28 \times 28$ image is normalized to the range $[0,1]$ and converted into a temporal spike sequence using Poisson rate coding across $T$ discrete time steps.

Feedforward SNNs with LIF neurons are implemented in PyTorch and snnTorch \cite{eshraghian2021training} to serve as software reference models. To evaluate temporal robustness, models are trained and tested with timestep windows of $T \in \{25, 50, 75, 100\}$ time steps, comparing cross-time configurations where training and inference durations differ. The networks employ hidden-layer sizes of 16, 32, 64, 128, and 256 neurons. For each network size, all $4 \times 4$ train/infer combinations are evaluated, yielding sixteen experiments per configuration and eighty experiments in total. This comprehensive exploration ensures temporal consistency and model generalization across various temporal resolutions.

RTL-level simulation is carried out using Synopsys VCS. A custom environment orchestrates the complete verification sequence: (i) configuration data and model parameters are loaded via dedicated memory initialization files; (ii) encoded spike stimuli are injected into the accelerator input buffers; and (iii) output spike activity is captured for decoding. Classification decisions are derived by integrating output spikes over the inference window, and the resulting labels are compared against ground-truth labels from the dataset to verify correctness.

\subsection{Power and Latency Analysis}

Power evaluation follows a multi-stage RTL-based methodology using Synopsys toolchains. Switching activity is captured from Synopsys VCS \cite{noauthor_vcs_nodate} simulations executing MNIST inference workloads. The generated activity files were combined with synthesized netlists produced by Synopsys RTL Architect \cite{noauthor_rtl_nodate}, targeting a 45\,nm CMOS OpenNand process library. Detailed power estimation was then performed using Synopsys PrimePower \cite{noauthor_primepower_nodate}, accounting for both dynamic and static components. This flow enables subsystem-level power attribution across neuron banks, synaptic weight units, spike forwarders, and on-chip memory structures.

The weight memory subsystem of Cerebra-H is implemented using dedicated SRAM macros generated with an open-source OpenRAM-based compiler \cite{noauthor_peramorphiqperamorphiq-sram-compiler_2026} and configured for the 45\,nm technology node. Each weight resolver memory block is organized as eight bit-sliced sub-banks, where each bank stores 128 bits per word across 2048 rows, enabling improved power distribution and compliance with SRAM compiler constraints. These SRAM macros were integrated into the synthesis and power analysis flow with corresponding timing and characterization views, ensuring accurate modeling of memory access latency and energy consumption within the overall estimation framework.

Latency and area characterization are performed using Synopsys RTL Architect. The analysis reports the maximum achievable operating frequency together with overall silicon area estimation of the design. Combined with power results, this evaluation provides a quantitative assessment of the operational efficiency of the Cerebra-H architecture under event-driven execution.

\section{Results}

While the proposed SNAP-V architecture encompasses a complete RISC-V-based SoC, the experimental evaluation presented in this section focuses specifically on characterizing the Cerebra-H neuromorphic accelerator. By isolating the accelerator's functional accuracy, power distribution, and architectural efficiency, these results demonstrate its strong potential to serve as a capable compute engine for resource-constrained edge environments.

\begin{table*}[t]
\caption{Comparison of Cerebra-H with representative neuromorphic systems including Area and Frequency metrics}
\label{tab:state_of_art_comparison_updated}
\centering
\resizebox{0.9\linewidth}{!}{%
\begin{tabular}{lcccccc}
\toprule
\textbf{Design} & \textbf{Tech Node} & \textbf{Area ($mm^2$)} & \textbf{Neuron Count} & \textbf{Freq. (MHz)} & \textbf{Power (W)} & \textbf{Energy (pJ/SOP)} \\
\midrule
\textbf{SNAP-V (This Work)} & \textbf{45 nm CMOS} & \textbf{25.74} & \textbf{1024} & \textbf{96.24} & \textbf{0.5001} & \textbf{1.05} \\
ODIN \cite{frenkel_0086-mm2_2019} & 28 nm FD-SOI & 0.086 & 256 & 75--100 & -- & 12.7 \\
OpenSpike \cite{modaresi_openspike_2023} & 130 nm & 33.3 & 1,024 & 40 & $\approx$ 0.225 & -- \\
TrueNorth \cite{akopyan_truenorth_2015} & 28 nm CMOS & 430 & 1,000,000 & 0.001 & 0.065 & 26.0 \\
Loihi 1 \cite{davies_loihi_2018} & 14 nm FinFET & 60 & 131,000 & - & - & 23.6 \\
Loihi 2 \cite{orchard_efficient_2021} & Intel 4 (7 nm) & 31 & 1,000,000 & $\approx$ 1000 & 1.55 & 10.1--11.5 \\
DYNAPs \cite{moradi_scalable_2018} & 180 nm CMOS & 43.79 & 1,024 & - & -- & 26.0 \\
SpiNNaker \cite{furber_spinnaker_2014} & 130 nm & 102 & 250,000+ & 200 & 1 & 2000-1000 \\
4096-Neuron SNN \cite{chen_4096-neuron_2019} & 10 nm FinFET & 1.72 & 4,096 & 105--506 & - & 3.8 \\
\bottomrule
\end{tabular}%
}
\end{table*}

\subsection{Functional Accuracy Evaluation}

The classification accuracy evaluation verifies the functional correctness of Cerebra-H across multiple SNN configurations. Table~\ref{tab:agg_accuracy_results} summarizes the aggregated results across all temporal settings and hidden layer sizes, covering a total of eighty experiments. Each configuration includes four training and four inference time-step conditions to provide a comprehensive assessment of temporal robustness.

\small
\begin{table}[t!]
\caption{Average functional accuracy across hidden layer sizes and all temporal configurations}
\label{tab:agg_accuracy_results}
\centering
\begin{tabular}{@{}>{\raggedright\arraybackslash}p{.22\columnwidth}
                >{\centering\arraybackslash}p{.14\columnwidth}
                >{\centering\arraybackslash}p{.15\columnwidth}
                >{\centering\arraybackslash}p{.15\columnwidth}
                >{\centering\arraybackslash}p{.13\columnwidth}@{}}
    \toprule
    \textbf{Hidden Layer} & \textbf{Software} & \textbf{Hardware} & \textbf{Diff.} & \textbf{Samples} \\
    \textbf{Neurons} & \textbf{Acc. (\%)} & \textbf{Acc. (\%)} & \textbf{(\%)} & \textbf{(Exp.)} \\
\midrule
16  & 92.83 & 87.11 & $-5.72$ & 16 \\
32  & 95.14 & 92.32 & $-2.82$ & 16 \\
64  & 96.45 & 93.94 & $-2.52$ & 16 \\
128 & 97.13 & 95.72 & $-1.42$ & 16 \\
256 & 97.32 & 96.69 & $-0.63$ & 16 \\
\midrule
    \textbf{Overall Avg.} & \textbf{95.77} & \textbf{93.16} & \textbf{$-2.62$} & \textbf{80} \\
\bottomrule
\end{tabular}
\end{table}
\normalsize

The results indicate an overall deviation of only 2.62\% between hardware and software inference accuracies. Notably, this deviation consistently decreases as the network size increases, indicating strong numerical stability in the Cerebra-H architecture for larger models. The 256-neuron configuration demonstrates the highest fidelity with a deviation of just 0.63\%, confirming that quantization and fixed-point arithmetic introduce minimal error for SNN workloads.

This trend suggests a statistical averaging effect in larger parameter spaces, where quantization errors are diluted. While the smaller configurations (e.g., 16 and 32 neurons) exhibit a higher relative deviation, this sensitivity is largely an artifact of the categorical classification task (MNIST) being evaluated, where such highly constrained networks lack representational redundancy. In contrast, small-scale SNN applications-such as the 20- to 40-neuron models used for robotic locomotion or PID control discussed earlier, typically process continuous dynamic signals. These control-oriented networks are inherently more robust to minor fixed-point quantization noise than categorical classifiers \cite{lohar_sound_2023, wang_stabilization_2016}. Consequently, with hardware accuracies remaining above 92\% even at 32 neurons, Cerebra-H maintains stable and accurate inferencing, demonstrating its strong potential for embedded neuromorphic tasks ranging from adaptive control to event-driven perception.

\subsection{System Level Power Characterization}

The component-level power distribution for the complete Cerebra-H is summarized in Table~\ref{tab:results_power_updated}. These results provide a detailed view of subsystem contributions to the total power consumption of 500.10\,mW. It should be noted that the reported total system power is dominated by the synthesized memory macros used in this technology node, rather than by the neuromorphic compute or communication logic.

\begin{table}[htbp]
    \centering
    \caption{Component level power breakdown for the complete Cerebra-H}
    \label{tab:results_power_updated}
    \resizebox{0.9\columnwidth}{!}{
    \begin{tabular}{lcc}
        \toprule
        \textbf{Subsystem} & \textbf{Power (mW)} & \textbf{Percentage (\%)} \\
        \midrule
        Weight Memory & 479.95 & 95.97 \\
        Neuron Clusters & 17.00 & 3.40 \\
        Spike Packet Paths & 2.44 & 0.49 \\
        Data/Control Packet Paths & 0.72 & 0.14 \\
        \midrule
        \textbf{Total System Power} & \textbf{\CerebraHTotPow} & \textbf{100.00} \\
        \bottomrule
    \end{tabular}
    }
\end{table}

A critical analysis of the power breakdown reveals a distinct dominance of storage overhead over computational cost. The active neuromorphic computation and communication, comprising the Neuron Clusters (3.40\%) and Spike Packet Paths (0.49\%), account for a combined total of less than 3.89\% of the accelerator's total power. In stark contrast, the Weight Memory consumes 95.97\% of the total power solely for synaptic weight storage and retrieval. This disparity demonstrates that the core logic and routing fabrics of Cerebra-H are highly optimized, yielding an average synaptic energy of 1.05\,pJ per synaptic operation (SOP) for the intrinsic compute pathway (excluding memory activity), while the system's total energy footprint is almost entirely dictated by the static and dynamic power costs of the memory units.

This observation highlights a critical design trade-off for future neuromorphic architectures. While distributing memory on-chip successfully mitigates the von Neumann bottleneck by localizing data access, the inherent power and area demands of this distributed storage can quickly overshadow the energy savings gained in the compute logic. Consequently, to fully realize the advantages of edge-oriented neuromorphic systems, distributing standard memory blocks is insufficient. The distributed memory infrastructure must be heavily optimized or adopt more biologically inspired distribution schemes. Addressing this dominant power sink implies that future architectures will heavily rely on integrating techniques such as synaptic weight compression, advanced hardware-aware mapping algorithms to maximize memory utilization, optimized cluster-level memory organization, and emerging paradigms like in-memory computing.

Timing characterization further reports a critical path delay of 10.3904\,ns, corresponding to a maximum operating frequency of 96.24\,MHz. Compared to the baseline Cerebra-S architecture, which operated at 10.17\,MHz, this significant acceleration directly validates the architectural enhancements of Cerebra-H. It specifically demonstrates how the hierarchical NoC, the localized compute parallelism from the clustered organization, and the concurrent weight memory access effectively eliminate prior communication bottlenecks. Crucially, this performance gain is achieved with minimal energy overhead, as the NoC consumes only 2.44\,mW (0.49\% of the total system power). This efficiency confirms that the hierarchical communication fabric delivers high throughput while strictly adhering to the power and resource constraints of small-scale edge systems. The total estimated silicon area of the design is 25.74\,mm$^2$ under the 45\,nm CMOS OpenNand library. These results define the upper bound on real-time throughput and provide a quantitative measure of the hardware footprint of the Cerebra-H architecture.

\subsection{Comparison with Existing Neuromorphic Systems}

To place Cerebra-H within the broader neuromorphic hardware landscape, a comparison was performed against representative digital neuromorphic systems. It is important to note that this comparison is not normalized for technology node, memory architecture, or memory implementation style, and reported power figures across platforms therefore reflect different system-level assumptions. Consequently, this comparison is intended primarily to provide broader architectural context rather than to rank performance directly. The evaluation focuses on technology node, neuron capacity, total power, and reported energy per synaptic operation, as summarized in Table~\ref{tab:state_of_art_comparison_updated}.

Cerebra-H achieving a low synaptic energy of 1.05\,pJ/SOP demonstrates that the accelerator's core event-driven pipeline, neuron update logic, and synapse accumulation units operate with high efficiency. When compared with the representative systems in Table~\ref{tab:state_of_art_comparison_updated}, this compute metric demonstrates highly competitive energy characteristics within its target scale of 1024 neurons. This confirms that the unique clustered organization, one-to-one mapping of logical to hardware neurons, and the low-power hierarchical NoC incur minimal overhead, enabling the architecture to perform highly efficiently compared to other designs.

Furthermore, with a total estimated silicon area of 25.74\,mm$^2$, Cerebra-H maintains a reasonable hardware footprint that places it favorably among representative digital architectures targeting embedded tasks. The operating frequency is also sufficiently high to ensure adequate timing margin for real-time spiking neural inference and low-latency execution. Together, these area, energy, and timing results position Cerebra-H as a viable, energy-efficient, and time-robust neuromorphic accelerator tailored for SNN inferencing.

Given its support for compact networks ranging from tens to a thousand neurons, the Cerebra-H accelerator integrated within the SNAP-V SoC presents strong architectural potential for application domains where existing large-scale neuromorphic platforms would be unnecessarily complex and resource-inefficient. While these applications remain targets requiring further system-level validation, the architectural characteristics and demonstrated performance metrics suggest that the SNAP-V platform holds strong potential for embedded control systems, event-based perception pipelines, and closed-loop robotic workloads operating at this scale.

\section{Conclusion}

This paper presented SNAP-V, a RISC-V-based System-on-Chip integrating a configurable neuromorphic accelerator (Cerebra-H) designed for energy-efficient inference of small-scale SNNs. By combining a distributed cluster-based organization with a hybrid NoC structure, the architecture addresses the computational and timing demands of event-driven processing while maintaining a flexible and tightly coupled interface with embedded RISC-V processors. The proposed design demonstrates the feasibility of integrated neuromorphic acceleration within a general-purpose SoC framework, offering both a potential solution for resource-constrained edge deployments and an accessible platform for neuromorphic research and prototyping.

Experimental validation confirms the robustness and numerical fidelity of the Cerebra-H architecture. Functional verification across eighty different configurations reveals strong agreement between hardware and software implementations, with an overall average inference accuracy deviation of only 2.62\%. Larger network configurations demonstrate even higher fidelity, reaching a minimal deviation of 0.63\%, confirming that fixed-point arithmetic and the event-driven execution pipeline preserve the numerical stability required for SNN inference. Timing analysis reports a maximum operating frequency of \CerebraHMaxFreq\,MHz in the 45\,nm CMOS OpenNAND process node, ensuring reliable real-time operation for embedded spiking workloads.

A key architectural insight from this work is the clear separation between computational efficiency and memory overhead. The accelerator achieves a highly competitive synaptic energy of 1.05\,pJ/SOP compared to the existing neuromophic systems. This reflects the intrinsic efficiency of the neuron update and communication network. However, the total system power of 0.5001\,W is overwhelmingly dominated by the weight memory subsystem, which accounts for approximately 95.97\% of total dynamic power consumption. This result highlights a fundamental system-level constraint: the overall energy profile is dictated primarily by synaptic weight storage and retrieval rather than by neuromorphic computation itself. SNAP-V serves both as a demonstration of highly efficient compute logic and as evidence that future gains in energy efficiency will depend heavily on memory architecture advancements, including optimized weight storage schemes, hierarchical buffering, and in memory computing. By establishing a verifiable, high-fidelity, and tightly integrated hardware platform, SNAP-V represents a meaningful step toward practical embedded neuromorphic computing and serves as a viable research platform for experimentation and validation.

\section*{Acknowledgment}
This work was supported by the Department of Computer Engineering, Faculty of Engineering, University of Peradeniya, Sri Lanka.

\section*{Author Contributions}
Kanishka Gunawardana, Sanka Peeris, and Kavishka Rambukwella contributed equally to the conceptualization, design, and implementation of the SNAP-V architecture and the Cerebra-H neuromorphic accelerator. Thamish Wanduragala and Saadia Jameel contributed equally to the conceptualization, design, and implementation of the Cerebra-S accelerator. Kanishka Gunawardana, Sanka Peeris, Kavishka Rambukwella, Thamish Wanduragala, and Saadia Jameel co-wrote the manuscript and contributed to the editing process. Roshan Ragel and Isuru Nawinne supervised the project, provided technical guidance, acquired funding, and provided overall research direction.

\section*{Declarations}
\begin{itemize}
    \item \textbf{Competing interests:} The authors declare no competing interests.
    
    \item \textbf{Data availability:} The MNIST dataset used for functional evaluation is publicly available and can be accessed via standard machine learning repositories (http://yann.lecun.com/exdb/mnist/). The trained Spiking Neural Network model weights (.pth files), the generated hardware deployment files, the synthesized netlists, power evaluation reports generated via Synopsys PrimePower, and latency characterization data supporting the findings of this study are available from the corresponding author upon reasonable request.
    
    \item \textbf{Code availability:} The source code for the SNAP-V SoC, implemented in Chisel using the Chipyard framework, along with the custom Verilog RTL for the Cerebra-S and Cerebra-H accelerators, the Python-based SNN training pipeline, and the custom hardware configuration compiler, are not publicly available due to ongoing research and development. However, the code and scripts are available from the corresponding author upon reasonable request and will be made available to editors and reviewers during the peer-review process.
\end{itemize}

\bibliographystyle{IEEEtran}
\bibliography{FYP}


\end{document}